\documentstyle[aps,preprint,tighten,floats,epsf,rotate,prl]{revtex}

\newbox\rotbox

\renewcommand{\baselinestretch}{1.1}

\begin{document}

\preprint{
\vbox{Submitted to {\it Physics Letters B}\null\hfill TRI-PP-95-47\\
        \null\hfill UW-PP-DOE/ER/40427-20-N95\\
         \null\hfill nucl-th/9507026
}}
 
\title{\LARGE $\rho-\omega$ Mixing via QCD Sum Rules\\ with\\ Finite
Mesonic Widths} 

\author{\sc M.J. Iqbal,$^{\rm a}$ Xuemin Jin,$^{\rm b}$\thanks{%
Address after September 1, 1996: Center for Theoretical Physics,
Laboratory for Nuclear Science and Department of Physics,
Massachusetts Institute of Technology, Cambridge, 
Massachusetts 02139} and 
Derek B. Leinweber$^{\rm b,c}$} 
\address{$^{\rm a}$Physics Department, The University of British
Columbia, Vancouver, B.C. V6T 1Z1, Canada}
\address{$^{\rm b}$TRIUMF, 4004 Wesbrook Mall, Vancouver, B.C. V6T
2A3, Canada} 
\address{$^{\rm c}$
Department of Physics, Box 351560, University of 
Washington, Seattle WA 98195, USA}
                                             
\maketitle

\begin{abstract}
  Based on the analysis of both Borel and Finite-Energy QCD sum rules,
  the inclusion of finite mesonic widths leads to a dramatic effect on
  the predictions for the momentum dependence of the $\rho-\omega$
  mixing matrix element.  It is shown that the $\rho-\omega$ mixing
  matrix element traditionally discussed in the literature, has the
  same sign and similar magnitude in the space-like region as its
  on-shell value.  This contrasts the zero-width result where the
  mixing matrix element is typically of opposite sign in the
  space-like region.
\end{abstract}
\pacs{PACS number(s):11.30.-j,13.75.Cs,14.40.Cs,11.55.Hx\\
{\it Keywords:}
Rho-omega mixing; Finite mesonic widths; QCD sum rule; Momentum
dependence; Dispersion relation}

\newpage
%%%%%%%%%%%%%%%%%%%%%%%%%%%%%%%%%%%%%%%%%%%%%%%%%%%%%%%%%%%%%%%%%%%%%

Recently, it has been argued that the $\rho-\omega$ mixing matrix
element should change significantly
off-shell\cite{goldman1,krein1,piekarewicz1,%
hatsuda1,mitchell1,connell1,urech1,mitra1,cohen1,maltman1,gao1}. 
This off-shell behavior would drastically reduce the contribution 
of $\rho-\omega$ mixing to the charge symmetry breaking (CSB) 
nucleon-nucleon potential\cite{goldman1,krein1,piekarewicz1,hatsuda1,%
mitchell1,connell1,urech1,mitra1,cohen1,maltman1,gao1,iqbal1}.
In a recent Letter\cite{iqbal2}, we have shown, in a model independent
way, that the finite widths of $\rho$ and $\omega$ mesons can give
rise to a new source of momentum dependence for the $\rho-\omega$
mixing matrix element.  The $q^2$ dependence arising due to the meson
widths leads to a significant alteration of the result obtained in the
zero-width approximation usually discussed in the literature. The
origin of this momentum dependence lies in the difference between the
$\rho$ and $\omega$ meson widths.  In Ref.~\cite{iqbal2} we have
quoted various values of the mixing parameter $\lambda$ extracted from
QCD sum-rule analysis that did not include mesonic
widths\cite{hatsuda1}. This Letter will explore the effect of finite
mesonic widths on the determination of $\lambda$.

To this end we invoke both the Borel QCD sum rules (BSR) and
Finite-Energy QCD sum rules (FESR). We will show that the large
difference between the $\rho$ and $\omega$ widths can have a dramatic
effect on the value of $\lambda$.  While it is positive in the zero
width limit, $\lambda$ becomes negative when the physical $\rho$ and
$\omega$ widths are included. Adopting this finite-width sum-rule
prediction for $\lambda$, we find that the $\rho-\omega$ mixing matrix
element in the space-like region, $-1.0\, {\rm GeV}^2\leq q^2\leq 0$,
has the same sign and a similar magnitude as its on-shell value.  
This contrasts the zero-width result where the mixing matrix element
is typically of opposite sign in the space-like region.

Consider the $\rho$ and $\omega$ mixed propagator defined by
\cite{hatsuda1}
\begin{equation}
\Pi^{\rho\omega}_{\mu \nu}(q)=i\int d^4x e^{iq\cdot x} \langle
0|T J^\rho_\mu (x) J^\omega_\nu(0)|0\rangle\ ,
\label{prop-def}
\end{equation}
where $J^\rho_\mu$ and $J^\omega_\nu$ are interpolating fields
representing the $\rho$ and $\omega$ mesons, respectively.  It has
been argued recently that there is no unique choice for these
interpolating fields~\cite{cohen1,maltman1}. Although this is
certainly the case, an obvious and physically preferred choice is the
standard vector currents. The reader is referred to Ref.~\cite{iqbal2}
for more discussions on this point.

It is common to find two factorizations of the Lorentz structure of
$\Pi^{\rho\omega}_{\mu \nu}(q^2)$ reflecting current conservation
\begin{eqnarray}
\Pi^{\rho\omega}_{\mu \nu}(q^2) &\equiv& - \left ( g_{\mu \nu} -
  {q_\mu \, q_\nu \over 
q^2} \right ) \, \Pi_1 (q^2) \, , 
\label{pi1}\\
\Pi^{\rho\omega}_{\mu \nu}(q^2) &\equiv& - \left ( q^2 g_{\mu \nu} -
  q_\mu \, q_\nu \right 
) \, \Pi_2 (q^2) \, ,% 
\label{pi2}% 
\end{eqnarray}%
from which the relation 
\begin{equation}
\Pi_1 (q^2) = q^2 \, \Pi_2(q^2) \, ,
\label{simple}
\end{equation}
follows trivially. 
There has been a great deal of interest
surrounding the value of $\Pi_1(q^2)$ at $q^2 =
0$~\cite{connell1,urech1,maltman1}.  The direct electromagnetic
contribution ($\rho\rightarrow\gamma\rightarrow\omega$) to the mixed
correlator is reasonably well understood and makes a numerically small
contribution to the correlator.  As such, this physics is usually
subtracted from the correlator~\cite{hatsuda1}, and the resulting
correlator focuses on the mixing of $\rho$ and $\omega$ mesons as well
as higher resonances.

It is then trivial to demonstrate that $\Pi_1(q^2)=0$ at $q^2 = 0$.
In the absence of intermediate state photons, the correlator $\Pi_{\mu
  \nu}(q^2)$ describes the propagation of physical hadrons.  As such,
$\Pi_{\mu \nu}(q^2)$ does not have any poles at $q^2 = 0$ as there are
no hadronic resonances below the threshold of $4 \, m_\pi^2$.  If
$\Pi_1 (q^2)$ is finite at $q^2 = 0$ then relation (\ref{simple})
indicates that $\Pi_2(q^2)$ has a pole at $q^2 = 0$.  This would be
perfectly acceptable provided there is a zero in the factor preceding
$\Pi_2(q^2)$ in (\ref{pi2}) at $q^2 = 0$ to cancel out the pole.  This
zero is absent in (\ref{pi2}).  Hence in order to correctly describe
the propagation of physical hadrons, $\Pi_2(q^2)$ cannot have a pole
at $q^2 = 0$.  From (\ref{simple}) it follows that $\Pi_1(q^2)=0$ at
$q^2 = 0$.  

The analytic structure of the mixed propagator allows us to
write a dispersion relation of the form,
\begin{equation}
{\rm Re\ } \Pi_1(q^2) = \frac{P}{\pi} \,
\int_{4 m_{\pi}^2}^{\infty}  \frac{{\rm Im}\, \Pi_1(s)
}{(s-q^2)} ds \, ,
\label{dis-gen}
\end{equation}
which is valid to leading order in quark masses with the neglect of
small electromagnetic interactions between quarks~\cite{hatsuda1}.  It
is clear that there are many excited states contributing to
$\Pi_1(q^2)$ in addition to the ground state $\rho$ and $\omega$
resonances. However, the $\rho-\omega$ mixing matrix element is 
traditionally defined by saturating the dispersion integral 
(\ref{dis-gen}) with  the ground state $\rho$ and $\omega$
resonances only~\cite{hatsuda1}
\begin{equation}
{\rm Re\ } \Pi_1 (q^2) \equiv  
\frac{\theta(q^2)}{(q^2-m_{\rho}^2)
(q^2 - m_{\omega}^2 )} \, .
\label{m-def-ha}
\end{equation}  
Consequently, the result, $\Pi_1(0) = 0$, is spoiled, and hence 
$\theta(0) \neq 0$. 

Saturating the dispersion integral of Eq.~(\ref{dis-gen}) by the
$\rho$ and $\omega$ resonances with the inclusion of finite mesonic
widths, one can express the $\rho-\omega$ mixing element defined above
as~\cite{iqbal2}
\begin{equation}
{\theta_{\Gamma} (q^2) \over \theta_\Gamma (m^2)} =  \Biggl \{ 
\frac{ G(q^2)}{G(m^2)}
\, \frac{ I_{\rho} (q^2) - I_{\omega} (q^2) }
{ I_{\rho} (m^2) - I_{\omega} (m^2) }
+\lambda \;
\frac{G(q^2)}{m^2}
\frac{ I_{\rho} (q^2)\,I_\omega (m^2) 
- I_{\omega} (q^2)\,I_\rho(m^2) }
{I_{\rho} (q^2) - I_{\omega} (q^2)}
\Biggr \}\ ,
\label{theta3f}
\end{equation}
where we have defined 
\begin{equation}
I_{\rho , \omega} (q^2) \equiv  {P\over \pi}
 \int_{4m_{\pi}^2}^{\infty} ds 
\frac{m_{\rho , \omega}\,\Gamma_{\rho ,
\omega}(s)}{(s-q^2) \left [ (s-\overline{m}_{\rho, \omega}^2)^2
+m_{\rho , \omega}^2 \, \Gamma_{\rho , \omega}^2) \right ] } \, ,
\label{def-in}
\end{equation}
\begin{equation}
G(q^2) \equiv  \frac
{ \left[(q^2 - \overline m_{\rho} ^2 )^2 + m_{\rho}^2 \,
\Gamma_{\rho}^2(q^2) \right]\, 
\left[ (q^2 - \overline m_{\omega} ^2 )^2 + m_{\omega}^2 \,
\Gamma_{\omega}^2(q^2) \right] } 
{ (q^2 - \overline m_{\rho} ^2)(q^2 - \overline{m}_{\omega}^2) -
m_{\rho} \, m_{\omega} \, 
\Gamma_{\rho}(q^2) \, \Gamma_{\omega}(q^2) } \, ,
\label{def-g}
\end{equation}
with $m^2\equiv (m_\omega^2+m_\rho^2)/2$,
$\overline{m}^2_{\rho,\omega} \equiv
m_{\rho,\omega}^2-\Gamma_{\rho,\omega}^2/4$, and
$\Gamma_{\rho,\omega}$ is the half width of the meson.  The momentum
dependence of the mesonic widths is indicated in equations
(\ref{def-in}) and (\ref{def-g}).  In particular, the widths vanish
below the threshold of $4\, m_\pi^2$.  For simplicity we shall
approximate $\Gamma_{\rho,\omega}(q^2)$ as constants given by their
physical values, $\Gamma_\rho = 151.5$ MeV and $\Gamma_\omega = 8.4$
MeV.  Inclusion of momentum dependent widths has a negligible effect
on the mixing in the space-like region of interest \cite{iqbal2}.

% $\Gamma_{\rho,\omega}$ is the half width of the meson. Note that
% the mesonic widths are in general momentum dependent. In particular,
% the widths vanish below the threshold. Here for simplicity we shall 
% approximate $\Gamma_{\rho,\omega}(q^2)$ as constants given by their physical 
% values, $\Gamma_\rho = 151.5$ MeV and $\Gamma_\omega = 8.4$ MeV. The
% results will be essentially identical when the momentum dependence of
% the widths are included.

% The momentum
% dependence of the mesonic widths is indicated in equations
% (\ref{def-in}) and (\ref{def-g}).  In particular, the widths vanish
% below the threshold of $4\, m_\pi^2$.  However, consideration of the
% uncertainties associated with the QCD sum rule analysis suggests that
% the use of $q^2$-independent widths will be sufficient for our
% numerical analysis.  The inspection of (\ref{def-g}) also reveals that
% the constant width terms are small relative to other terms in the
% space-like regime. In the time-like region, the effects of the
% momentum dependence of the mesonic widths only have a marginal
% smoothing effect at the threshold (see Ref.~\cite{iqbal2}).

Our task now is to determine the mixing parameter, $\lambda$, 
using the QCD sum-rule approach.
QCD sum rules for $\rho-\omega$ mixing are obtained by studying the
correlation function of (\ref{prop-def}) with
$J^\rho_\mu=(\overline{u}\gamma_\mu u-\overline{d}\gamma_\mu d)/2$ and
$J^\omega_\nu=(\overline{u}\gamma_\mu u+\overline{d}\gamma_\mu d)/6$.
The sum rules of $\Pi^{\rho\omega}_{\mu\nu}(q)=\left(q_\mu q_\nu
  -g_{\mu\nu}q^2\right)\Pi^{\rho\omega}(q^2)$ are derived for
$12\,\Pi^{\rho\omega}(q^2)$ using a dispersion relation. The BSR can
be written as\cite{svz,hatsuda1}
\begin{equation}
{12\over \pi}\int_{4 m_{\pi}^2}^{\infty} {\rm Im}\,\Pi^{\rho
  \omega}(s) 
\, {e^{-s/ M^2}}  ds =c_{0}\,M^2+c_1+{c_2\over M^2}+{c_3\over M^4}\ ,
\label{borel-sumrule}
\end{equation}
where $M$ is the Borel mass and $c_0$, $c_1$, $c_2$ and $c_3$ have
been given in the literature.
\begin{eqnarray}
& &c_0={\alpha_e\over  16\pi^3}\ , \hspace*{4.7cm}
c_1={3\over 2\pi^2}\left(m_d^2-m_u^2\right)\sim 0\ ,
\nonumber
\\*[7.2pt]
& &c_2=4 {m_u-m_d (1+\gamma)\over 2+\gamma}\langle\overline{q}q
\rangle_0\ , \hspace*{1.5cm}
c_3={224\pi\,\gamma\over 81}\alpha_s \kappa\,\langle\overline{q}q
\rangle_0^2
-{28\pi\over 81}\alpha_e \kappa\,\langle\overline{q}q
\rangle_0^2\ ,
\label{coef}
\end{eqnarray}
where $\gamma\equiv \langle\overline{d}d
\rangle_0/\langle\overline{u}u \rangle_0-1$, $\langle\overline{q}q
\rangle_0\equiv (\langle\overline{u}u \rangle_0+\langle\overline{d}d
\rangle_0)/2$, and $\alpha_e$ and $\alpha_s$ are the electromagnetic
and strong coupling constants, respectively.  Here we have truncated
the operator product expansion (OPE) at dimension six and kept only
the terms considered in the literature.

The studies of $\rho-\omega$ mixing via the QCD sum-rule approach have
been made by Shifman, Vainshtein, and Zakharov~\cite{svz}, and
subsequently by Hatsuda {\it et al.}\cite{hatsuda1}. In these works,
the mesonic widths were ignored, and the
spectral density ${\rm Im}\, \Pi^{\rho \omega}(s)$ has been
characterized by poles at the masses of the $\rho$, $\omega$,
$\rho^\prime$ and $\omega^\prime$ mesons plus a continuum model
starting at an effective threshold $s_0$\cite{svz,hatsuda1}:
\begin{eqnarray}
{12\over \pi}{\rm Im}\, \Pi^{\rho \omega}(s)
=& &f_\rho\,\delta (s-m_\rho^2)-f_\omega\,\delta (s-m_\omega^2)
+f_{\rho^\prime}\,\delta (s-m_{\rho^\prime}^2)
\nonumber
\\*[4.2pt]
& &
-f_{\omega^\prime}\,\delta (s-m_{\omega^\prime}^2)
+c_0\,\theta(s-s_0)\ .
\label{spectral-0}
\end{eqnarray}
While the assumption of poles for lowest resonances may be reasonable
for the determination of resonance parameters, here we are concerned
with subtle cancelations of the nearby $\rho$ and $\omega$ strengths.
To include the effect of finite mesonic widths, we replace the
$\delta$ functions in Eq.~(\ref{spectral-0}) by normalized
Briet-Wigner distributions for the meson widths of the form
\begin{equation}
\frac{1}{\pi}\, \frac{m_{\rm v} \, \Gamma_{\rm v}}
  {(s-\overline m^2_{\rm v})^2  + m^2_{\rm v} \, \Gamma^2_{\rm v}} \,
  . 
\end{equation}
The BSR can then be expressed as
\begin{eqnarray}
f_{\rho} \, \tilde{I}_{\rho}(&M^2&) - f_{\omega} \,
\tilde{I}_{\omega}(M^2) \, + 
f_{\rho^\prime} \, \tilde{I}_{\rho^\prime}(M^2)
- f_{\omega^\prime} \, \tilde{I}_{\omega^\prime}(M^2) 
\nonumber
\\*[7.4pt]
&=&c_0\,M^2\,\left[ 1 - e^{-s_0/M^2} \right] + c_1
+ {c_2\over M^2} + {c_3\over M^4}\ ,
\label{sumrule-widths}
\end{eqnarray}
where we have defined 
\begin{eqnarray}
  \tilde{I}_{\rm v}(M^2) &\equiv& \frac{1}{\pi}
     \int_{4m_{\pi}^2}^{\infty}  ds \,
     \frac{m_{\rm v}\,\Gamma_{\rm v}\, e^{-s/M^2} }{
    (s-\overline{m}_{\rm v}^2)^2 +m_{\rm v}^2 \, \Gamma_{\rm v}^2}\ .
\label{i-finite-widths}
\end{eqnarray}
The quantity of interest, $\lambda$, is related to $f_\rho$ and
$f_\omega$ through
\begin{equation}
\lambda = \frac{g_{\rho} \, g_{\omega}}{12} \left[ \frac{f_{\rho}}
{m_{\omega}^{2}} - \frac{f_{\omega}}{m_{\rho}^{2}} \right] \,
 \frac{m^2}{\Theta (m^2)}\ ,
\label{lambda-cal}
\end{equation}
where $\Theta (m^2)$ is the on-shell value of the $\rho-\omega$ mixing
with the electromagnetic contribution subtracted. Following the
earlier works we take $g_{\rho}^{2}/4\pi = 2.4$, $g_{\omega} = 3 \,
g_{\rho}$, and $\Theta (m^2) = -0.0051 {\rm GeV}^2$.

In Ref.\cite{hatsuda1}, the authors used the BSR of
(\ref{borel-sumrule}), the first derivative sum rule as obtained by
taking the derivative of (\ref{borel-sumrule}) with respect to
$1/M^2$, a FESR, and the on-shell constraint in their analysis. 
While the absence of strong
continuum model contributions makes the $\rho-\omega$ mixing
correlator a reasonable candidate for FESR analysis, we note that
consideration of (\ref{sumrule-widths}) alone is sufficient to extract
the parameters of interest. This latter approach sidesteps the serious
problems associated with derivative sum rules\cite{jin1,leinweber1}.

The on-shell constraint given in Refs.~\cite{svz,hatsuda1} is valid
for the zero-width case only. For the finite-width case it becomes
\begin{equation}
f_{\omega} = \frac{m_{\rho}^{2}}{m_{\omega}^{2}}\,
\frac{ {\rm I}_{\rho}(m^2)}{ {\rm I}_{\omega}(m^2)}\, f_\rho
+ \frac{12}{g_{\rho} \, g_{\omega}}\,\frac{m_{\rho}^2}{
{\rm I}_{\omega}(m^2)}\,\frac{\Theta (m^2)}{G(m^2)}\ .
\label{constraint}
\end{equation}

The BSR analysis is carried out following the techniques for
determining the valid Borel region as introduced in Ref.\ 
\cite{leinweber2}.  We limit the continuum model contributions to
50\%, and maintain the contributions of the highest dimensional
condensates in the operator product expansion (OPE) to less than 10\%
of the sum of terms in the OPE.  The valid Borel regime throughout
this work is $1.15 \le M \le 2.45$ GeV.

Uncertainties in the extracted parameters are determined via the Monte
Carlo based uncertainty analysis introduced in Ref.\
\cite{leinweber1}; uncertainties on QCD parameters such as condensate
values are utilized to provide uncertainties in the OPE.  Here we use
$-4\pi^2\langle\overline{q}q\rangle_0=0.62\pm 0.05{\rm GeV}^3$,
$\kappa=2\pm 1$ and $\alpha_s/\pi (1 {\rm GeV}^2)=0.117\pm 0.014$
\cite{leinweber1}.  The average quark mass is chosen to satisfy the
Gell-Mann--Oakes--Renner relation. The two new parameters encountered
in (\ref{borel-sumrule}) include $m_u / m_d = 0.50 \pm 0.25$ and
$\gamma = 0.008 \pm 0.002$.  
The uncertainty in the quark mass ratio is taken from the particle
data tables \cite{PDT} and is much larger than the uncertainty of $\pm
0.04$ assumed in Ref.\ \cite{hatsuda1}. In turn, these OPE
uncertainties provide distributions for the extracted parameters, from
which uncertainty estimates are obtained.  Since the resultant
distributions are not Gaussian, we use the more robust measure of the
median for the central value, and extract standard error estimates
away from the median.

\vspace{12pt}
\noindent{\small\bf 1. Zero-Width Case}

To begin our analysis, we fix the continuum threshold, $s_0$, at the
$\rho(1700)$ resonance mass and optimize the fit in this region by
adjusting the strength of the $\rho^\prime$ and $\omega^\prime$
contributions.  We utilize the constraint of (\ref{constraint}), and
extract $\lambda$ via (\ref{lambda-cal}).  Table \ref{table-1}
summarizes the optimal fit parameters.  These results are in
reasonable agreement with Ref.\ \cite{hatsuda1}.

\begin{table}[t]
\renewcommand{\baselinestretch}{1.3333333}
\caption{Borel sum rule predictions for the $\rho - \omega$ mixing
  parameters.} 
\setdec 0.000000
\begin{tabular}{lcccc}
Parameter    &\multicolumn{2}{c}{Zero Widths} 
&\multicolumn{2}{c}{Physical Widths} \\
                        &            &         
&Constrained    &No Constraint \\
\tableline
$f_\rho \times 10^2$    &$1.91 {+0.02 \atop -0.01}$   &$1.89 {+0.03
  \atop -0.01}$ 
&$0.303{+0.052 \atop -0.026}$  
&$0.28{+0.27 \atop -0.15}$  \\
$f_\omega \times 10^2$  &$1.95{+0.02 \atop -0.01}$
                        &$1.97 {+0.01 \atop -0.02}$
&$0.2488{+0.0004 \atop -0.0008}$  
&$0.23{+0.20 \atop -0.11}$  \\
$f_\phi \times 10^2$    &$--$            &$0.07\pm0.04$  
&$-0.04\pm0.04$  
&$-0.04\pm0.05$  \\
$f_{\rho\prime} \times 10^2$   &$-0.11\pm0.20$  &$--$  &$--$  &$--$ \\
$f_{\omega^\prime} \times 10^2$ &$-0.15\pm0.23$  &$--$  &$--$  &$--$\\
$s_0$ (GeV$^2$)             &$2.89$          &$3.20$  
&$6.00$  
&$5.52$  \\
$\lambda$                &$1.63 {+0.34 \atop -0.55}$   &$2.11 {+0.37
  \atop -0.61}$   
&$-0.64{+0.38 \atop -0.75}$  &$-0.62{+0.48 \atop -0.93}$  \\
%
%\tablenotemark[1] 
%
\end{tabular}
\label{table-1}
\end{table}

Of course, the $\phi$ meson should be included in the parameterization
of the spectral density \cite{maltman1}.  To keep the same number of
parameters in the fit we replace the individual contributions of
$\rho'$ and $\omega'$ by an effective strength $f_{\rho'\omega'}$ at
the average $\rho'$ and $\omega'$ mass.  Optimization of the three
parameters $f_\rho$, $f_\phi$ and $f_{\rho'\omega'}$ sends the
strength from the $\rho'$, $\omega'$ region to zero.  Hence, we
discard the parameter $f_{\rho'\omega'}$ in favor of optimizing the
continuum threshold.  Strength from this region is very small and has
little effect on the optimal value of $\lambda$.

The continuum threshold is not precisely determined by the sum rule.
We find $\sqrt{s_0}=1.79\pm 1.50\,{\rm GeV}$.  This is due to a lack
of excited state strength in the sum rules as governed by the leading
terms of the OPE.  To avoid the consideration of unphysical values for
$s_0$, we optimize $s_0$ for the central values of the condensates,
and then fix it for the Monte-Carlo sampling of the QCD parameters
throughout the following.  We note that fixing $s_0$ to the optimal
value, as opposed to keeping this parameter as a search parameter in
the uncertainty analysis, has a negligible effect on the central
values and uncertainties of the other fit parameters.

\vspace{12pt}
\noindent{\small\bf 2. Physical-Width Case}

For the finite physical-width case, we first consider a fit of
$f_\rho$, $f_\phi$ and $s_0$ and determine $f_\omega$ from the
constraint of (\ref{constraint}).  The central result of this paper is
that $\lambda$ changes sign due to the large finite width of the
$\rho$ meson. 
% The $\phi$ meson makes a negligible contribution.  
The value of $s_0$ is larger than that for the zero width case,
indicating a reduction of strength in the excited state regime,
including the $\rho^\prime$, $\omega^\prime$ region.

Using the on-shell constraint effectively fixes one of the parameters
we are trying to extract from the QCD sum rules.  Since there could be
a discrepancy between the on-shell value and the QCD parameters we
have selected, it is imperative to relax this constraint.  Since the
$\rho$-meson width is much larger than the $\omega$-meson width, there
is now a considerable difference between their contributions to the
sum rules.  Thus, it is possible to resolve the individual
contributions of $\rho$ and $\omega$ mesons without the external
constraint.  The ``No Constraint'' column of Table \ref{table-1}
summarizes the results of this four parameter fit.  While the absolute
values of $f_\rho$ and $f_\omega$ are not as well determined, their
correlations lead to a value for $\lambda$ that is essentially
unchanged from the constrained case and with only slightly larger
uncertainty.  The sum rule prediction for the on-shell value is
$\Theta(m^2) = -0.0046 {+0.0022 \atop -0.0040}\, {\rm GeV}^2$ which
agrees with the physical value of $-0.0051\, {\rm GeV}^2$.

\begin{figure}[p]
\begin{center}
\epsfysize=11.6truecm
\leavevmode
\epsfbox{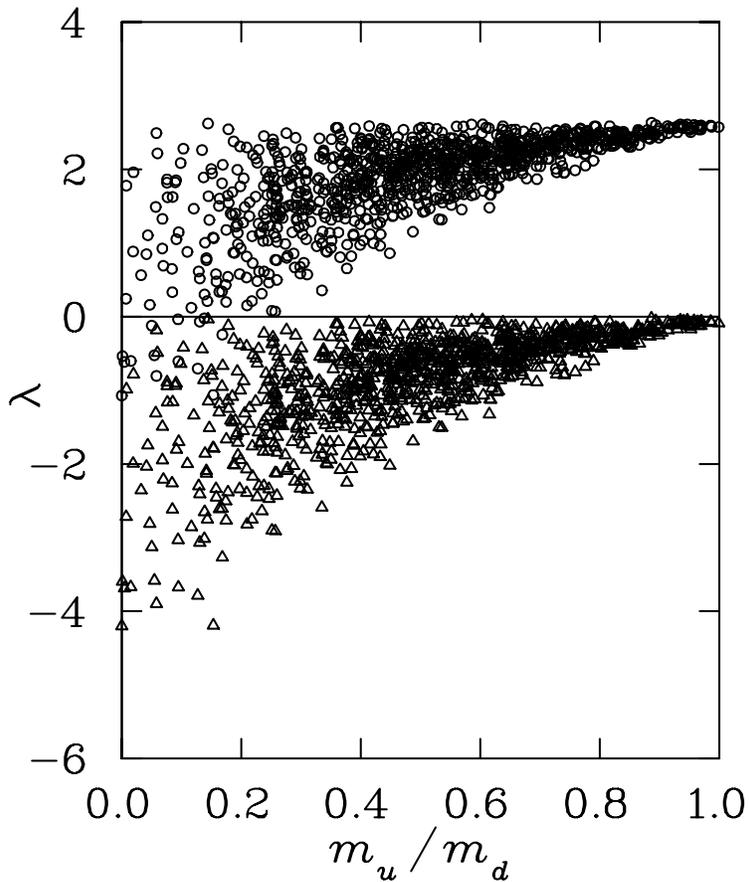}
\end{center}
\caption{Scatter plot of $\lambda$ obtained from $1000$ QCD parameter
  samples. The correlation with the quark mass ratio is displayed. The
  effect of including finite mesonic widths ($\triangle$) is to shift
  down the zero-width results ($\circ$) and produce negative values
  for $\lambda$.}
\label{fig-1}
\end{figure}

In searching for correlations between the QCD parameters of the OPE
and the $\rho - \omega$ mixing parameter $\lambda$, only one
interesting correlation arises.  Fig.~\ref{fig-1} displays a scatter
plot for $\lambda$ and the quark mass ratio $m_u/m_d$ where the
constraint equation of (\ref{constraint}) is used.  If the constraint
equation is relaxed for the finite width case, the distribution
remains largely unchanged. Finally, if we adopt the value for
$m_u/m_d=0.57\pm 0.04$ estimated in \cite{xpt} and used in
\cite{hatsuda1}, we find $\lambda=-0.52\pm 0.32$, using the constraint
of (\ref{constraint}).

\begin{table}[p]
\renewcommand{\baselinestretch}{1.33333}
\caption{Finite-Energy sum rule predictions for $\rho - \omega$ mixing
  parameters.} 
\label{borel}
\setdec 0.000000
\begin{tabular}{lcccc}
Parameter    &\multicolumn{2}{c}{Zero Widths} 
&\multicolumn{2}{c}{Finite Widths} \\
                        &          &       
&Physical Widths  &Equal Widths\\
\tableline
$f_\rho \times 10^2$    &$1.87$   &$1.85$
&$0.304$  
& $1.86$ \\
$f_\omega \times 10^2$  &$1.91$
                        &$1.93$
&$0.248$  
&$1.93$  \\
$f_\phi \times 10^2$    &$--$            &$-0.081$  
&$0.021$  
&$-0.073$  \\
$f_{\rho^\prime}\times 10^2$   &$-0.17$  &$--$  &$--$  &$--$  \\
$f_{\omega^\prime} \times 10^2$ &$-0.21$  &$--$  &$--$  &$--$  \\
$f_{\rho^\prime\omega^\prime}\times 10^2$   &$--$  &$-0.0003$
&$-0.0037$  &$0.0016$  \\ 
$s_0$ (GeV$^2$)             &$3.0$          &$3.0$  
&$3.0$  
&$3.0$  \\
$\lambda$                &$1.53$   &$2.13$  
&$-0.67$  &$2.02$  \\
%
%\tablenotemark[1] 
%
\end{tabular}
\label{table-2}
\end{table}

Since the mixed correlator has little excited state strength, we have
also performed the FESR analysis. Results are shown in
Table~\ref{table-2}. We see that the FESR results are in good
agreement with the BSR results. To emphasize that the change in sign
of $\lambda$ occurs due to the large difference of $\rho$ and $\omega$
widths, we show the results for a hypothetical case of finite but
equal meson widths in column 5. In this case the results agree with
the zero-width case.

\begin{figure}[t]
\begin{center}
\epsfysize=11.6truecm
\leavevmode
\setbox\rotbox=\vbox{\epsfbox{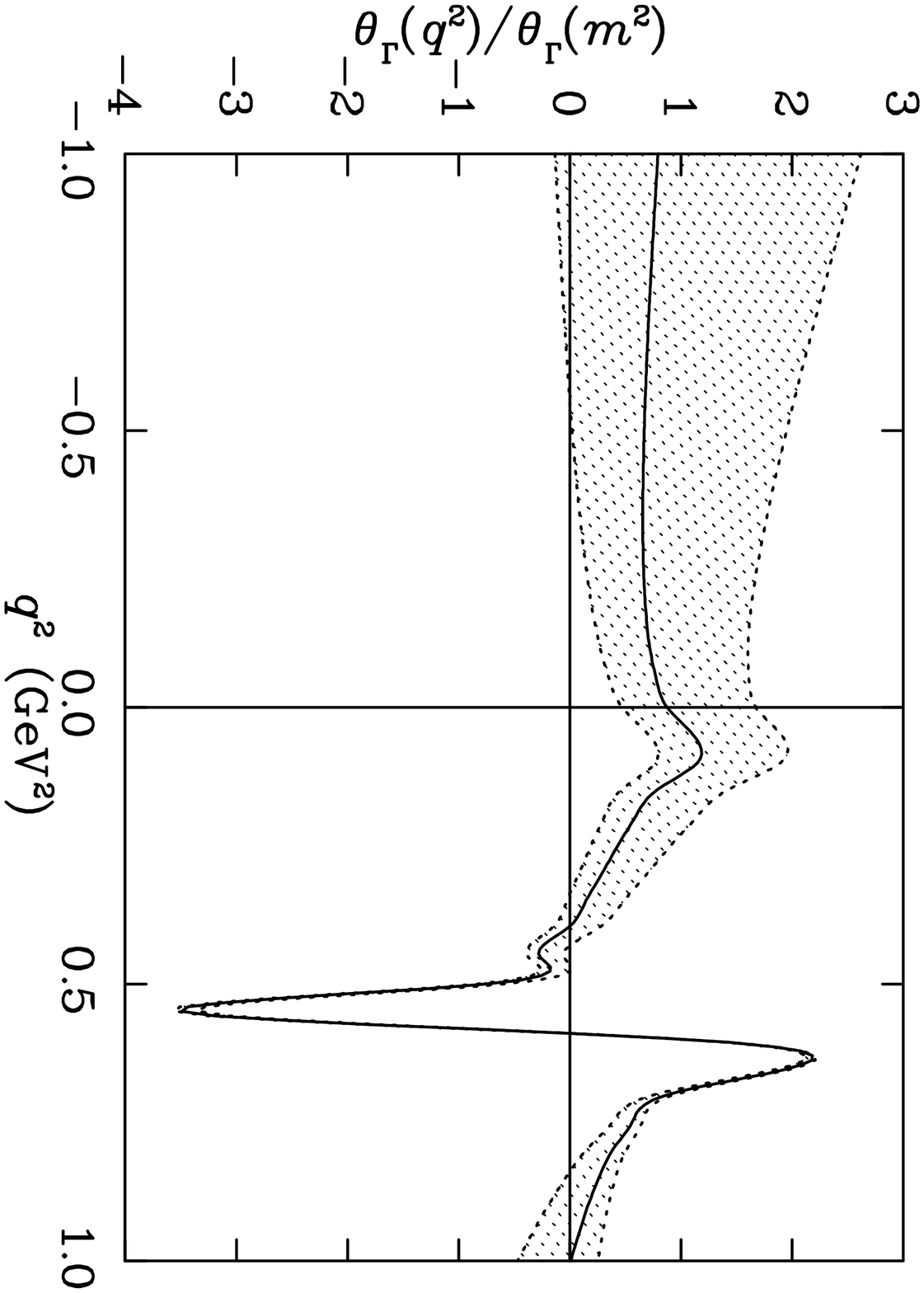}}\rotl\rotbox
\end{center}
\caption{$\theta_{\Gamma}(q^2)/\theta_\Gamma(m^2)$ as a function of
  $q^2$ for $\lambda=-0.64 {+0.38 \atop -0.75}$ obtained from BSR
  analysis with the on-shell constraint. The solid curve corresponds
  to the median value, and the two dotted curves illustrate the
  standard error. The two spikes in the time-like region correspond
  to the two poles of $G(q^2)$ [see Eq.~(\protect{\ref{def-g}})].}
\label{fig-2}
\end{figure}

The implications of finite mesonic widths on the $q^2$ dependence of
$\theta_\Gamma(q^2)/\theta_\Gamma(m^2)$ are illustrated in
Fig.~\ref{fig-2}. The ratio $\theta_\Gamma(q^2)/\theta_\Gamma(m^2)$ is
positive in the space-like $q^2$ region in contrast to the zero-width
results.  For the median value of $\lambda$, the $\rho-\omega$ mixing
matrix element in the space-like region has the same sign and similar
magnitude as its on-shell value.
A comprehensive analysis of the impact
of the present result on the CSB potential and corresponding physical
observables will be documented in a future publication \cite{iqbal3}.

In the present paper, the $\rho-\omega$ mixing matrix element is
defined to include only the ground state $\rho$ and $\omega$.  The
contributions of excited states may also be included in a more
complete definition.  However, this will require the knowledge of the
couplings of the interpolating fields to the excited states.  The
feasibility of reliably determining such couplings via QCD sum rules
is currently under investigation.

\vspace{7pt} 

We would like to thank Ernie Henley, Jerry Miller, and Richard
Woloshyn for helpful discussions and comments.  This work was
supported by the Natural Sciences and Engineering Research Council of
Canada and the US Department of Energy under grant DE-FG06-88ER40427.
%%%%%%%%%%%%%%%%%%%%%%%%%%%%%%%%%%%%%%%%%%%%%%%%%%%%%%%%%%%%%%%%%%%%%%
 
%\renewcommand{\baselinestretch}{1.0}

\end{document}